\documentclass[ApJL,twocolumn]{aastex62}
\usepackage{amssymb, amsmath, graphicx, dcolumn, units, xspace, mathtools, physics, tensor, bm, lipsum, revsymb}
\usepackage[normalem]{ulem}
\usepackage{gensymb}
\usepackage{xcolor}
\usepackage{subfigure}

\newcommand{\kms}{ km\,s^{-1} }

\newcommand{\vlos}{ v_\text{los} }

\newcommand{\SPA}{School of Physics and Astronomy, Monash University, Clayton VIC 3800, Australia}
\newcommand{\OzGravMonash}{OzGrav: The ARC Centre of Excellence for Gravitational Wave Discovery, Clayton VIC 3800, Australia}

\shorttitle{Meet the parents}
\shortauthors{Paynter \textit{et al}.}

\begin{document}

\title{Meet the parents: the progenitor binary for the supermassive black hole candidate in E1821+643}

\author{James Paynter}
\affiliation{School of Physics, University of Melbourne, Parkville, Victoria, 3010, Australia}

\author{Eric Thrane}
\affiliation{\SPA}
\affiliation{\OzGravMonash}

\date{\today}

\begin{abstract}
The remnants of binary black hole mergers can be given recoil kick velocities up to $\unit[5,000]{km\,s^{-1}}$ due to anisotropic emission of gravitational waves.
E1821+643 is a recoiling supermassive black hole candidate with spectroscopically offset, broad emission lines, consistent with motion of the black hole at $\sim\unit[2100]{km\, s^{-1}}$ along the line of sight relative to its host galaxy.
This suggests a recoil kick of $\sim \unit[2,200]{km\,s^{-1}}$.
Such a kick is powerful enough to eject E1821+643 from its $M_\text{gal}\sim2 \times \unit[10^{12}]{M_\odot}$ host galaxy.
In this work, we address the question: assuming that E1821+643 is a recoiling black hole, what are the likely properties of the progenitor binary that formed E1821+643?
Using astrophysically motivated priors, we infer that E1821+643 was likely formed from a binary black hole system with masses of $m_1\sim 1.9^{+0.5}_{-0.4}\times \unit[10^9]{M_\odot}$, $m_2\sim 8.1^{+3.9}_{-3.2} \times \unit[10^8]{M_\odot}$ (90\% credible intervals).
Given our model, the black holes in this binary were likely to be spinning rapidly with dimensionless spin magnitudes of ${\chi}_1 = 0.87^{+0.11}_{-0.26}$, ${\chi}_2 = 0.77^{+0.19}_{-0.37}$.
Such a high recoil velocity is impossible for spins aligned to the orbital angular momentum axis.
This suggest that the progenitor for E1821+643 merged in hot gas, which is thought to provide an environment where spin alignment from accretion proceeds slowly relative to the merger timescale.
We infer that E1821+643, if it is a recoiling black hole, is likely to be rapidly rotating with dimensionless spin ${\chi} = 0.92\pm0.04$.
A $\unit[2.6\times10^9]{M_\odot}$ black hole, recoiling from a gas-rich environment at $v\sim \unit[2,200]{km\,s^{-1}}$ is likely to persist as an active galactic nuclei for $\sim \unit[10^7]{yr}$, in which time it traverses $\sim \unit[25]{kpc}$.
\end{abstract}

\section{Introduction}\label{sec:intro}
Supermassive black hole (SMBH) binaries form during galaxy mergers~\citep{Begelman1980Natur.287..307B}.
These binaries can be driven to merge through a variety of processes including scattering from bulge stars \citep{Sesana2006}, interactions with the circumbinary disk \citep{Haiman2009}, and hardening from subsequent galaxy mergers \citep{Ryu2018}.
The asymmetric emission of gravitational radiation during coalescence carries linear momentum, giving the nascent black hole a kick~\cite{Bekenstein1973ApJ}.
Breakthroughs in numerical relativity revealed that merging black holes can achieve kick velocities up to $\sim \unit[5,000]{km\,s^{-1}}$~\cite{Baker2006ApJ...653L..93B,2006PhRvL..96k1101C,2007PhRvL..98w1102C,2011PhRvL.107w1102L}.
Recoiling SMBH which have been ejected from the galactic core, but not the galaxy itself, then experience a period of damped simple harmonic oscillations for $\sim \unit[1]{Gyr}$~\citep{Gualandris2008ApJ...678..780G,BlechaLoeb2008}.
However, a sufficiently large kick can completely eject the SMBH from its host galaxy, especially at early times when haloes were much less massive~\cite{Baker2006ApJ...653L..93B}.
A SMBH completely ejected from its host galaxy will exist as an AGN until it exhausts its accretion disk.
It will then wander the extremely low-density intergalactic medium, visible only through the rare lensing of background sources.

E1821+643 is an extremely luminous quasar at redshift $z=0.297$.
The central supermassive black hole candidate is estimated to have a mass of $\sim2.6\times\unit[10^9]{M_\odot}$ in a host galaxy with $M_\text{gal}\sim 2\times \unit[10^{12}]{M_\odot}$~\citep{Floyd2004MNRAS,Shapovalova2016ApJS}.
There is a $90\degree$ kink in the south-west radio jet~\citep{Blundell1996ApJ}, likely due to a spin-flip of the central SMBH engine during a recent merger~\citep{MerrittEkers2002Sci}.
Spectropolarimetric analysis has found the broad-line region is moving away from the observer relative to the host galaxy~\citep{Robinson2010ApJE1821,Jadhav2021MNRAS}.
The broad-line region is associated with the large-velocity-dispersion gas directly surrounding the SMBH.
Thus, E1821+643 is interpreted to be exiting the host galaxy at a speed of $2070\pm \unit[{50}]{km\, s^{-1}}$ along the line-of-sight~\citep{Robinson2010ApJE1821}.
If E1821+643 is indeed recoiling a black hole, then the merger which created the kink in the radio jet is almost certainly the same merger which ejected the quasar from the galactic nucleus.

While E1821+643 is not universally accepted as a recoiling supermassive black hole, there are three features of this object, which lead us to regard the recoiling black hole hypothesis as probable: the spatial offset of E1821+643 from the center of the host galaxy photocenter, the large spectral offset from the host galaxy, and a $90^\circ$ kink in the associated radio jet consistent with a spin flip of the central engine~\citep{MerrittEkers2002Sci}.

\cite{Jadhav2021MNRAS} simulate a spectro-astrometry observation based on the Hubble Space Telescope O III image to test whether the observed displacement is actually caused by spatial asymmetry in the O III emission as opposed to an offset quasar nucleus. 
They find that only one third of the observed displacement can be attributed to the asymmetric O III emission, suggesting the quasar is significantly offset from the host galaxy photocenter. 
Naively, one could attribute the large spectral offset not to a recoil kick, but to binary motion about a second supermassive black hole \citep{Jadhav2021MNRAS}. 
However, the implied binary orbital period of $\unit[\sim 10^4]{years}$ is too large to explain the $\unit[\sim1-10]{year}$ periodicity observed by~\cite{Shapovalova2016ApJS,kovacevic2017Ap&SS.362...31K,kovacevic2018MNRAS.475.2051K}.
While we cannot rule out the binary hypothesis definitively, quasars are known to exhibit quasiperiodic oscillations that are sometimes mistaken for binarity \citep{LiuT18,Vaughan16_false,SuperBayes,smbbh_min,Witt2022}, and so the presence of quasiperiodic oscillations are most simply explained in terms of red noise in the accretion disk.

In this Letter, we use Bayesian inference to estimate the properties of the progenitor binary that formed E1821+643 under the assumption that E1821+643 is a recoiling black hole.
The remainder of this work is organized as follows.
In Section~\ref{sec:method} we describe our methodology.
In Section~\ref{sec:models} we describe three different models for the merger environment, which determine the prior distributions for the binary parameters.
In Section~\ref{sec:results}, we present the results of our inference calculation and discuss the implications.

\section{Method}\label{sec:method}
Our goal is to calculate the posterior distribution $p(\theta|d)$ for SMBH progenitor binary parameters $\theta$ given the line-of-sight velocity data for E1821+643.
This posterior is given by Bayes theorem 
\begin{align}\label{eq:Bayes}
    p(\theta|d) \propto {\cal L}(d|\theta) \pi(\theta | M) ,
\end{align}
where ${\cal L}(d|\theta)$ is the likelihood function for the data $d$ given binary parameters $\theta$ and $\pi(\theta|M)$ is the prior given model $M$.
At the outset, we do not know ${\cal L}(d|\theta)$, but we can calculate it from the ``raw-data'' likelihood function describing the original line-of-sight velocity measurement ${\cal L}(d|v_\text{los})$ along with a binary black hole kick calculator.

We assume that the raw-data likelihood function is Gaussian:
\begin{align}
    {\cal L}(d | v_\text{los} ) .
\end{align}
The mean is given by the data $d=\unit[2070]{km\,s^{-1}}$, which is the measured line-of-sight velocity from \cite{Robinson2010ApJE1821} while the width is given by the associated error bar $\unit[50]{km\, s^{-1}}$.
The raw-data likelihood is conditioned on the unknown true line-of-sight velocity $v_\text{los}$.
The next step is to convert the raw-data likelihood to a likelihood for the data given the \textit{kick velocity} $v_k$, which takes into account the unknown transverse motion of the black hole.
The kick velocity is related to the inclination angle $\iota$ through the following equation:
\begin{equation}
    \vlos = v_k \cos\iota .
\end{equation}
We obtain our new likelihood by marginalising over the unknown inclination angle
\begin{equation}
    {\cal L} (d | v_k) = \int d(\cos\iota) \, \pi(\cos\iota ) {\cal L} (d | v_k \cos\iota) ,
\end{equation}
using a uniform prior prior for $\cos\iota$.
Marginalising over $\cos\iota$ causes the most likely kick velocity to be larger than the most likely line-of-sight velocity.

Even so, this new likelihood function underestimates the true kick velocity because we do not take into account gravitational acceleration from the host galaxy, which slows the black hole by an unknown amount in between the moment of birth and its observation in the present day. 
From~\citep[Fig.~7]{BlechaLoeb2008} we see that a billion-solar-mass black hole will escape its host galaxy with $v_k > v_\text{esc} \sim \unit[1,500]{km\, s^{-1}}$ for an ejection in any direction.
Since the line-of-sight velocity of E1821+643 is greater than the escape velocity of the galaxy, we expect that it is destined to escape.
The kick velocity is too large to exhibit the damped harmonic oscillations explored by~\cite{Gualandris2008ApJ...678..780G,BlechaLoeb2008}.

The next ingredient is the prior distribution for kick velocity,
\begin{align}
    \pi(v_k | \theta) ,
\end{align}
which is conditioned on the binary parameters $\theta$.
We are primarily interested in how the distribution of kick velocities depends on the the black hole mass ratio $q\equiv m_2/m_1$ and the dimensionless black hole spin magnitudes $\chi_1, \chi_2$.
If we fix these three parameters, the distribution of kick velocities $\pi(v_k|\theta)$ is broadened when we marginalize over the other relevant degrees of freedom---the orientation of the black-hole spin vectors $\theta_1, \theta_2$ and the phase of coalescence $\phi_c$---which can strongly affect $v_k$.
We calculate $\pi(v_k|\theta)$ using the gravitational-wave kick calculator \textsc{Precession}~\citep{precession}.

In order to make sure our estimates of $\pi(v_k|\theta)$ are sufficiently accurate, we test how our results change when we vary the initial orbital separation from $\sim\unit[0.1]{pc}$, i.e., from $r\sim 500,000 \, r_g$ to $r\sim \, 10 r_g$.
Here, $r_g = GM/c^2$ is the gravitational radius of the premerger binary mass $M=M_1+M_2$.
We find that the change in $\pi(v_k|\theta)$ is for the most part, negligible.
Only the phase of coalescence $\phi_c$ distribution exhibits a noticeable change.

With these ingredients, it is straightforward to calculate the likelihood of the data given the binary parameters
\begin{align}
    {\cal L}(d | \theta) = \int dv_k \, 
    {\cal L}(d | v_k) \pi(v_k | \theta) .
\end{align}
Now that we have the likelihood function needed for Eq.~\ref{eq:Bayes}, we turn attention to the final ingredient: $\pi(\theta|M)$ the prior distribution of the binary parameters given some model $M$.
Choosing this prior is tantamount to describing a model for the environment in which the SMBH binary merges.
We discuss these models in the next section.

\section{Models}\label{sec:models}
\begin{figure*}
    \centering
    \includegraphics[width=\textwidth]{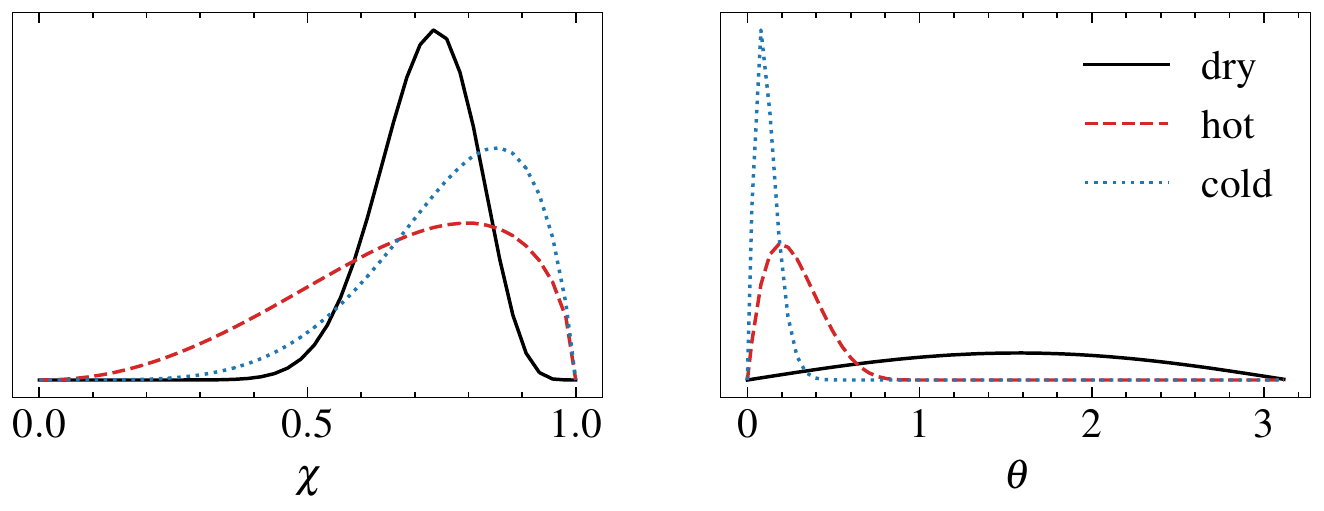}
    \caption{
    The distribution for dimensionless black-hole spin $\chi$ (left) and the tilt angle $\theta$ between the black-hole spin vector and the orbital angular momentum of the supermassive binary black hole.
    Different models for the merger environment are denoted with different colors.
    }
    \label{fig:models}
\end{figure*}

During galaxy mergers, the supermassive black holes quickly sink to the new galactic core, surrounded by dense gas. 
The corotation of the two supermassive black holes in the bulk accretion disk forms a gravitationally-bound binary system. We consider a class of models developed by \cite{Dotti2010MNRAS.402..682D,Lousto2012PhysRevD}, who consider a supermassive binary black hole embedded in a massive circumnuclear disk of gas surrounded by a larger stellar spheroid.  Following~\cite{Dotti2010MNRAS.402..682D,Lousto2012PhysRevD} we employ three models for the interaction between the SMBH binary and surrounding gas: dry, hot, and cold.
In each model, which are depicted graphically in Fig.~\ref{fig:models}, the gas equation of state affects how efficiently the gas accretion is able to align the black hole spins to the orbital plane~\cite{Bogdanovic2007ApJ}.
This in turn affects the kick velocity distribution.

It is important to note that these models are not based on precise calculations. 
They provide plausible sketches to describe the environments in which SMBH binaries merge, which is useful for gaining a qualitative understanding of objects like E1821+643. The properties of the progenitor binary for E1821+643, derived below using these models, are of course subject to unknown theoretical uncertainties.

\paragraph{Dry merger model.}
The dry merger model assumes there is minimal gas accretion onto the binary before merger.
For the dry model, we take uniform priors on the spin tilt angles with respect to the orbital plane.
The priors on $\theta_{1,2}$ are uniform on $\cos{\theta}\in[-1,1]$, i.e., there is no accretion alignment before merger.
\cite{LoustoZlochower2014PhRvD} find that a population of repeatedly merging black holes converges to a steady-state solution with ${\chi}\sim 0.7$.
We fit a beta distribution to Fig.~23 of \cite{LoustoZlochower2014PhRvD}, finding parameters $\alpha_{\chi,d}=17$, $\beta_{\chi,d} = 6.7$.
We use this beta distribution as the prior on both $\chi_1$ and $\chi_2$ for the dry model.

\paragraph{Hot merger model.}
The hot merger model assumes that the gas surrounding the binary pair is monatomic, with polytropic gas index $\gamma = 5/3$.
For the hot merger model, we adopt the beta distributions suggested by \cite{Lousto2012PhysRevD}, with $\alpha_{\chi,h}=3.2$, $\beta_{\chi,h} =1.6$ for a hot disk.
Similarly, the inclination angle $\theta$ are defined through parameters $\alpha_{\theta,h}=2.0$, $\beta_{\theta,h} =5.2$.
These priors truncate $\theta$ on $(0,1)$ rather than $(0,\pi)$, which they justify as there is minimal prior support in either the hot or cold scenario for $\theta\gtrsim 0.8$.
A comparison of these fits to the simulation results is given in \cite{Lousto2012PhysRevD}.

\paragraph{Cold merger model.}
The cold merger model approximates solar metallicity gas, with $\gamma = 7/5$.
The progenitor spin distribution follows a beta distribution with and $\alpha_{\theta,c}=5.9$, $\beta_{\theta,c} =1.9$ for cool gas.
The priors on inclination angle are more sharply aligned to the orbital angular momentum axis, $\alpha_{\theta,c}=2.5$, $\beta_{\theta,c} =20$.

\begin{figure*}
    \centering
    \begin{subfigure}
    \centering
    \includegraphics[width=\columnwidth]{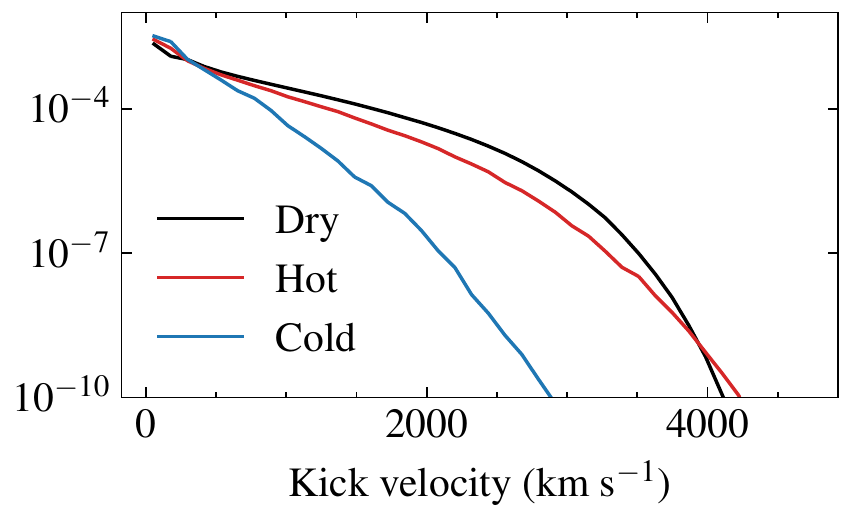}
    \label{fig:kickvelocities}
     \end{subfigure}
     \hfill
     \begin{subfigure}
    \centering
    \includegraphics[width=\columnwidth]{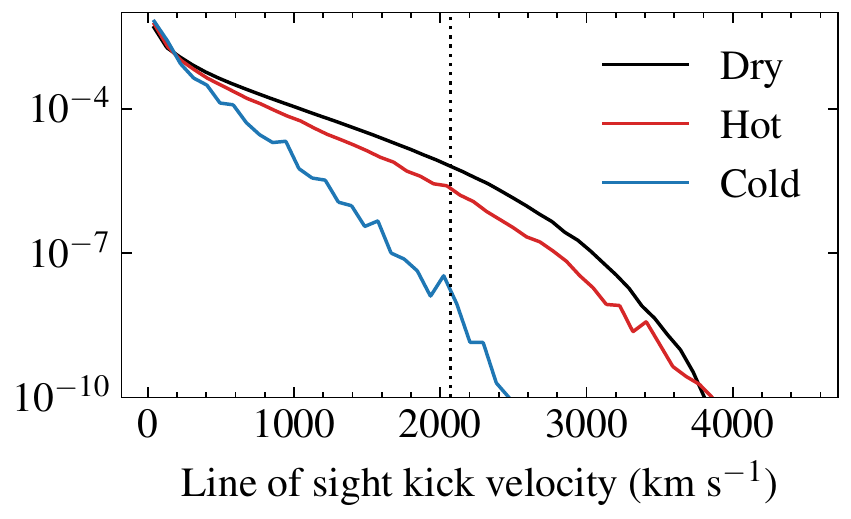}
    \label{fig:loskicks}
    \end{subfigure}
    \caption{
    Left: the distribution of kick velocity $v_k$ for the three models considered Section~\ref{sec:models} describing the environment of the supermassive binary black hole merger that formed E1821+643.
    Right: the distribution of line-of-sight velocity $v_\text{los}$ for the same three models.
    }
    \label{fig:velocities}
\end{figure*}

The progenitor mass ratio is not strongly affected by accretion, and is consistent across the three models.
Following~\cite{Lousto2012PhysRevD} our prior for mass ratio is
\begin{equation}
    \pi(q) \propto q^{-0.3}(1-q) .
\end{equation}
In Fig.~\ref{fig:velocities}, we show the distribution of kick velocity $v_k$ and line-of-sight velocity $v_\text{los}$ for each model.
Large line-of-sight velocities greater than or equal to the $v_\text{los}\approx\unit[2000]{km\, s^{-1}}$ observed for E1821+643 are unusual for all three of the models considered, occurring in just $\lesssim 0.23\%, 0.068\%, 0.00039\%$ of SMBH mergers in dry, hot, and cold mergers respectively. 
However, they are relatively far more common in dry and hot environments relative to cold environments.\footnote{One may wonder if the fact that E1821+643 exhibits an unusual recoil velocity in all of our model variants suggests that the models are qualitatively wrong. In statistical parlance, this is akin to asking if the models are ``misspecified.'' While black-hole remnants consistent with E1821+643 occur with a frequency of about one in $400$ for the dry model, we do not regard this as evidence for misspecification. After all, many SMBH candidates are known to science, and researchers do not write papers about all of them---only the most interesting ones. Selection effects may be important as well since $\unit[200]{km \, s^{-1}}$ kicks are challenging to disentangle from the normal stellar velocity dispersion. If, however, subsequent studies reveal that a large fraction of SMBH merger candidates recoil with unusually large velocities, this could be a sign of misspecification.
} 
Kick velocity probabilities are tabulated in Table~\ref{tab:vtable}, and Table~\ref{tab:lostable} for line-of-sight kick velocities. While kicks $v_k \gtrsim \unit[2000]{\kms}$ are rare, kicks of several hundred kilometres per second are common. In fact, it is extremely difficult to keep the recoiling AGN in the galactic nucleus, as $p(v_k>\unit[100]{\kms})>0.5$ under all models for pre-merger gas accretion.

\begin{table*}[]
    \centering
    \begin{tabular}{l r r r r r r }
          & P$(v_k>100)$ & P$(v_k>500)$ & P$(v_k>1,000)$ & P$(v_k>2,000)$ & P$(v_k>3,000)$ & P$(v_k>4,000)$ \\
        Dry & 0.76 & 0.36 & 0.15 &        $1.8\times 10 ^{-2}$ &        $4.0\times 10 ^{-4}$ &        $3.1\times 10 ^{-8}$ \\
        Hot & $0.72$ &       $0.25$ &        $8.1\times 10 ^{-2}$ &        $5.7\times 10 ^{-3}$ &        $8.5\times 10 ^{-5}$ &        $7.4\times 10 ^{-8}$ \\
        Cold &       0.69 &       0.11 &        $9.7\times 10 ^{-3}$ &        $2.9\times 10 ^{-5}$ &        $1.5\times 10 ^{-9}$ &        $1.7 \times 10^{-16}$ \\
    \end{tabular}
    \caption{Table of recoil velocity probabilities. Velocities are denoted in $\unit[]{km \, s^{-1}}$. 
    }
    \label{tab:vtable}
\end{table*}

\begin{table*}[]
    \centering
    \begin{tabular}{l r r r r r r }
          & P$(v_k>100)$ & P$(v_k>500)$ & P$(v_k>1,000)$ & P$(v_k>2,000)$ & P$(v_k>3,000)$ & P$(v_k>4,000)$ \\
        Dry &  0.55 &   0.16 &        $4.3\times 10 ^{-2}$ &        $2.3\times 10 ^{-3}$ &        $1.9\times 10 ^{-5}$ &        $2.1\times 10^{-10}$ \\
        Hot & 0.45 & $9.4\times 10 ^{-2}$ & $2.0\times 10 ^{-2}$ &        $6.8\times 10 ^{-4}$ &        $4.0\times 10 ^{-6}$ &   $1.7\times 10 ^{-9}$ \\
        Cold & 0.37 &  $2.7\times 10 ^{-2}$ & $1.3\times 10 ^{-3}$ &        $3.9\times 10 ^{-6}$ &        $8.3\times 10 ^{-12}$ &        $4.9\times 10 ^{-23}$ \\
    \end{tabular}
    \caption{Table of line-of-sight recoil velocity probabilities. Velocities are denoted in $\unit[]{km \, s^{-1}}$.}
    \label{tab:lostable}
\end{table*}

\section{Results \& Discussion}\label{sec:results}
We find the dry model is the most preferred by the data, followed closely by the hot model ($\ln$ Bayes factor $-2.0$).
The cold model is strongly disfavored with $\ln$ Bayes factor $\sim 7.6$ relative to the dry model.
(A $\ln$ Bayes factor of 8 is considered strong evidence in favour of one model over another~\citep{intro}.)
Since E1821+643 is observed as a quasar, we know it must carry a significant amount of gas with it from the merger process.
This is only possible in the hot or cold scenario, so we immediately rule out the dry merger model.
Henceforth all results presented are using the hot merger prior; it provides an adequate explanation of the large line-of-sight velocity for E1821+643 while accounting for the quasar nature of the host galaxy.

\begin{figure*}
    \centering
    \includegraphics[width=2.1\columnwidth]{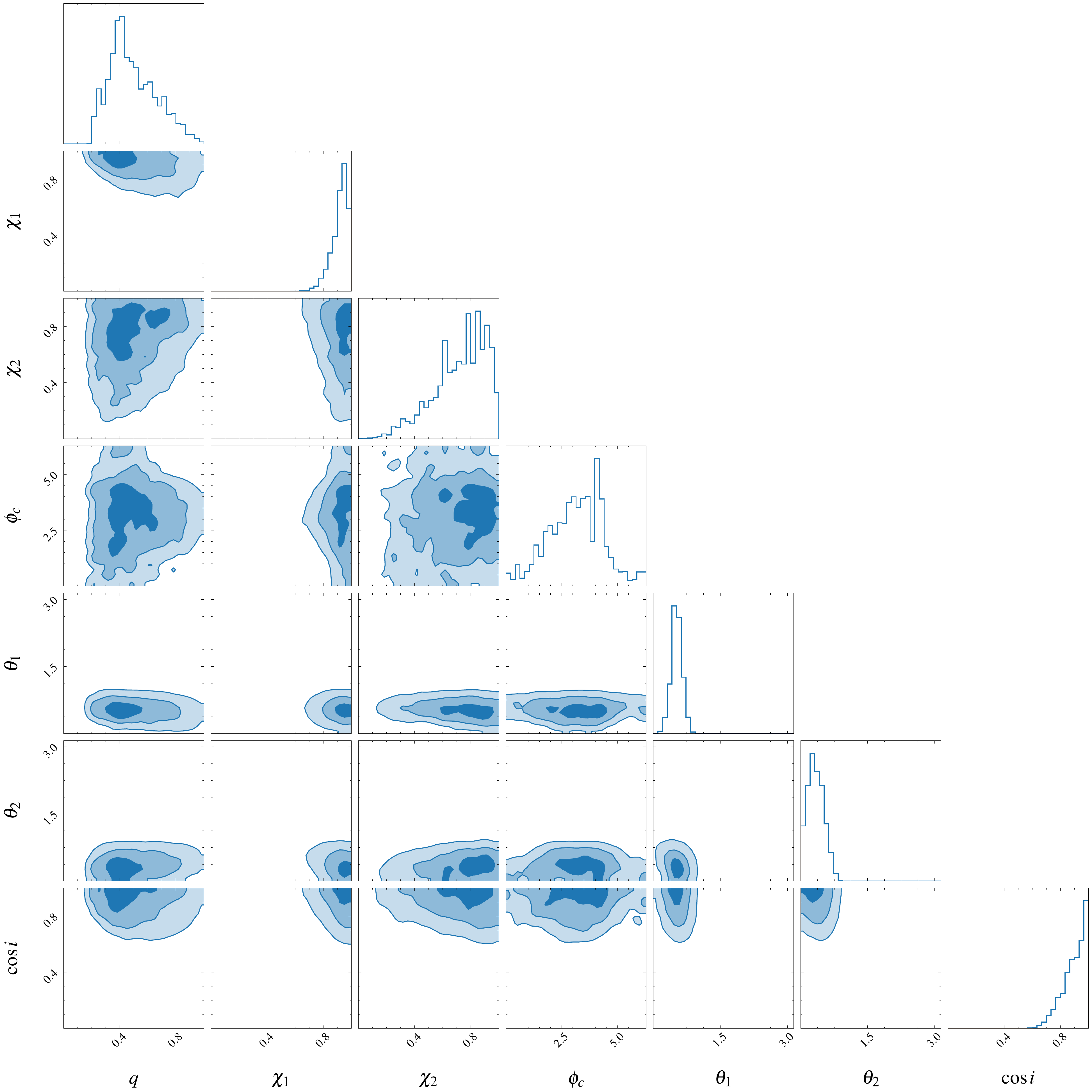}
    \caption{
    A corner plot showing the posterior distribution for parameters describing the progenitor binary for E1821+643 for the best-explanation ``hot'' model.
    The priors are indicated by dashed curves while the solid curves represent the posterior distribution.
    }
    \label{fig:corner}
\end{figure*}

A corner plot, showing the posterior distribution for our model parameters is provided in Fig.~\ref{fig:corner}.
Using the hot merger prior, we estimate the recoil velocity to be $v_k = \unit[2,200^{+480}_{-200}]{km \, s^{-1}}$, not accounting for velocity lost climbing out of the galactic potential to date {  ($90\%$ credibility)}.
We estimate the dimensionless spin to be ${\chi} = 0.92\pm0.04$ (90\% credibility).
We find the mass ratio to be $q=0.51 ^{+0.33}_{-0.24}$ { (90\% credibility)}.
We find the supermassive binary black hole progenitor of E1821+643 had masses of $m_1\sim 1.9^{+0.5}_{-0.4}\times \unit[10^9]{M_\odot}$, $m_2\sim 8.1^{+3.9}_{-3.2} \times \unit[10^8]{M_\odot}$ (90\% credible intervals).\footnote{The kick velocity is entirely determined by the mass ratio and black hole spin vectors, and so our estimates for these quantities are insensitive to uncertainty about the mass of E1821+643.
If the mass of E1821+643, which we take to be $\unit[2.6 \pm 0.3 \times 10^9]{M_\odot}$ \citep{Floyd2004MNRAS,Shapovalova2016ApJS}, were off by a factor of $\kappa$, then our estimates for component mass would be off by the same factor of $\kappa$. 
However, this would not affect our conclusion about the probable environment in which E1821+643 was formed, since it is based on the black hole spin vectors.
}
The black holes in this binary were likely to have dimensionless spin magnitudes of ${\chi}_1 = 0.87^{+0.11}_{-0.26}$, ${\chi}_2 = 0.77^{+0.19}_{-0.37}$.\footnote{It may at first seem remarkable that these binary parameters can be so tightly constrained. However, it is not so surprising when we consider the fact that only a small fraction of the binary parameter space is able to produce a kick large enough to explain the observed velocity of E1821+643.}

It is interesting to compare our results to those obtained through X-ray observations of E1821+643.
X-ray observations can be used to probe black-hole spin as follows.
The innermost stable circular orbit (ISCO) of a black hole depends on the black hole's spin.
The ISCO varies monotonically with $\chi$ from $R_\textsc{ISCO}= 6 r_g$ (for $\chi=0$) to  $R_\textsc{ISCO}= 1.23 r_g$ (for $\chi=0.998$).
The ISCO in turn sets the inner edge of the accretion disk, such that the redshifts measured from spectral lines of the ISCO should indicate the (gravitational + Doppler) redshifts of the accreting plasma.
The spectrum may either be dominated by thermal emission from the disk, or be predominantly illuminated by the X-ray corona which sits above (and below) the black hole along the spin axis.

Using measurements of the iron Fe-K$\alpha$ line at $\unit[6.4]{keV}$ (rest frame energy), \cite{2022MNRAS.tmp.2359S} find features in the X-ray spectrum of E1821+643 in excess of the power-law continuum characteristic of direct coronal emission.
By making the assumption that these features are caused by relativistic reflection from the inner edge of the accretion disk at the ISCO, they find E1821+643 has dimensionless spin $\chi = 0.62^{+0.22}_{-0.37}$ (90\% credible interval)~\citep{2022MNRAS.tmp.2359S}.

While the maximum-posterior values for $\chi$ from these two studies differ by $\Delta\chi = 0.30$, the results are broadly consistent given the error bars.
There is an $\gtrsim 11\%$ chance of getting such discrepant results due to noise fluctuations---if the two analyses had used the same priors.
In reality, \cite{2022MNRAS.tmp.2359S} assume a uniform prior in $\chi$ while our prior favours larger values of $\chi$ since we assume that E1821+643 was created from a merger event.
Taking into account the different choice of priors, the two measurements agree even better.
It is remarkable that the spin of E1821+643 can be estimated through such different means: X-ray emission from the accretion disk and inferences from its gravitational-wave recoil kick.

It is worth considering if there are other models, beyond the three we employ here, that might provide a plausible explanation of the data.
Another possible pre-merger scenario is chaotic accretion~\citep{2006MNRAS.373L..90K,2007MNRAS.377L..25K,2008MNRAS.385.1621K}.
In this scenario, black holes can grow rapidly through deposition of material from random directions.
This chaotic accretion of angular momenta keeps the black hole spin low, which allows the black hole to rapidly accrete material without violating the Eddington limit. Chaotic accretion may explain how $10^9-10^{10} M_\odot$ are possible at $z=6-10$ without requiring primordial intermediate-mass black-hole seeds.
Furthermore, chaotic accretion may solve the final parsec problem for supermassive  black hole binaries, ensuring that the binary merges within a Hubble time.
However, it has been shown by \cite{2013MNRAS.434.1946N} that chaotic accretion onto a \textit{binary} is likely to dramatically speed up accretion and thus alignment of the black hole spins. 
In a gas-rich merger where the amount of gas is not a limiting factor, the accretion rate can be increased up to $10^4$ times through shearing of the accretion disk~\citep{2013MNRAS.434.1946N}.
\cite{2013ApJ...774...43M} show that the black hole spins can align with the binary orbital axis $10-100$ times faster than the  orbital axis aligns with the accreting gas.
Both rates are rapid with respect to the binary evolution timescales until gravitational-wave emission dominates the system evolution. 
They find that this alignment occurs even when the gas arrives in small packets from uncorrelated directions; i.e., under chaotic accretion.
Therefore we suggest chaotic accretion is a poor explanation for the large observed line of sight kick velocity of the putative recoiling black hole.

A SMBH completely ejected from its host galaxy is expected to survive as an AGN until it burns through the remainder of its accretion disk~\citep{Loeb2007PhRvL,2008ApJ...687L..57V}. 
The amount of gas remaining bound to a moderately fast recoiling black hole is approximately a few percent of the black hole mass~\citep{2016MNRAS.456..961B}. 
A recoiling SMBH ejected at the host galaxies escape velocity will persist as an AGN for $\unit[\sim 10^7]{years}$, in which time it will traverse $\sim \unit[25]{kpc}$~\citep{2016MNRAS.456..961B}.

The merger that produced the recoiling AGN in E1821+643 likely occurred quite recently (in relative terms), as it is barely $\unit[1]{kpc}$ from the galactic core.
If we assume constant velocity, then the merger occurred roughly $\unit[420,000]{yr}$ ago.
This is consistent with the analysis of \cite{Robinson2010ApJE1821}, who estimate the merger and subsequent kick to have occurred $\sim \unit[10^5]{years}$ ago based on the kink in the radio jet $\sim \unit[1]{arcsec}$ from the jet head.
The jet kink implies that the jet source underwent a merger-induced spin flip~\cite{MerrittEkers2002Sci}.

Another explanation for the 90 degree kink in the jet include precession of a premerger binary~\citep{Blundell1996ApJ,Blundell2001ApJ...562L...5B}, though we regard this is unlikely due to the spatial and spectral line velocity offset of E1821+643. 
Alternatively, the kink may be due to jet interactions with the intergalactic medium~\citep{2004ApJ...609..539K}. \cite{2013AJ....146..120L} track 200 AGN over 17 years, finding significant bending and jet reorientation in 60 of the most heavily observed objects. 
In general, kinks can be quite shallow, but appear as right-angles due to projection effects~\citep{2004ApJ...609..539K}.

\cite{Jadhav2021MNRAS} find that the SMBH is $\sim \unit[580]{pc}$ southeast of the galactic core.
Accounting for line-of-sight effects ($\cos i \approx 27\degree$), we find the SMBH has moved $\sim \unit[1.5]{kpc}$ since the merger.
\cite{Floyd2004MNRAS} estimate the mass of the host galaxy to be $M\sim 2\times \unit[10^{12}]{M_\odot}$, with a half-light radius of $\unit[18.9]{kpc}$, fitting a de~Vaucouleurs profile~\citep{deVaucouleurs1959}.
The change in velocity of the black hole climbing from $\sim \unit[1]{pc}$ to $\sim \unit[1.5]{kpc}$ is negligible.

In the future, it may be possible to study the population properties of supermassive black holes using surveys of black-hole line-of-sight velocities.
With a sufficiently large catalog of measurements, it might be possible to say something about the mass and spin distribution of merging supermassive black holes based on the kick velocities of their recoiling remnants.

It is interesting to consider what any putative recoiling billion-solar-mass black holes would look like today.
These ``hypercompact stellar systems'' (HCSS) would have gone dark long ago, having accreted their disks~\cite{OLearyLoeb2009MNRAS,MerrittSchnittmanKomossa2009ApJ...699.1690M}.
The cluster of stars they carry with them would have almost certainly evolved into stellar remnants---black holes, white dwarfs, and neutron stars---making these HCSS hard to detect.
Limits exist~\citep{2021MNRAS.507L...6C} on the number density of such objects in the Universe, but they are not very constraining, as rapidly recoiling, billion-solar-mass black holes are expected to be rare.
The lensing rate of HCSS on background sources is low, again due to their relative scarcity in the universe.
These supermassive black holes will lens their cluster of stars, which may generate observable signals~\cite{RauchBlandford1994ApJ}, Paynter et al. (in prep).

\begin{acknowledgments}
E.T. acknowledges support from the Australian Research Council (ARC) Centre of Excellence CE170100004 and DP230103088.
The authors acknowledge the feedback of the anonymous referee who made numerous helpful suggestions to improve this manuscript.
\end{acknowledgments}

\bibliography{refs}

\end{document}